\documentclass[12pt]{iopart}
\usepackage{graphicx} 
\usepackage[english]{babel}
\usepackage[utf8]{inputenc}
\usepackage{xcolor}

\usepackage{gensymb}
\begin{document}

\title{Simple physics behind the flight of a drone} \author{Martín
  Monteiro$^{1,2}$, Cecilia Stari$^2$, Cecilia Cabeza$^2$, Arturo
  C. Mart{\'i}$^2$ }

\address{$^1$ Universidad ORT Uruguay}
\address{$^2$ Instituto de F\'{i}sica, 
  Universidad de la Rep\'{u}blica,
 Uruguay}

\ead{marti@fisica.edu.uy}

\date{\today}
\begin{abstract}
The flight of a quadcopter drone, readily available as a toy, is
analyzed using simple physics concepts. A smartphone with built-in
accelerometer and gyroscope was attached to the drone to register the
accelerations and angular velocities along the three spatial axis
while the drone is taking off, landing or rotating.  The vertical
speed, the height and one of the angular coordinates are obtained
through numerical integration of the acceleration values and compared
with information provided by the manufacturer. The analysis of these
quantities provides an opportunity to gain insight into important
physics concepts involving Newton laws and conservation principles in
a stimulating environment.
\end{abstract}

\maketitle

\section{Drones and Physics}

Remotely-controlled helicopters and planes have been available as toys
for decades. However, only recently, advances in sensor technologies
have made possible to easily fly and control theses devices at an
affordable price \cite{floreano2015science}. Along with their
increasing availability the educational opportunities are also
proliferating \cite{Obaid_2021,saifizi2021using}.  In a previous
experiment we measured the height and pressure in the inner layer of
the atmosphere using a smartphone mounted on a drone. These kinds of
measurements can be compared with approximate models as an isothermal
or a constant density atmosphere and also with the International
Standard Atmosphere \cite{monteiro2016exploring}.  Here, a simple
experiment in which a smartphone was mounted on a drone is described,
which investigates the basics of flight.  Thanks to the smartphone's
built-in accelerometer and gyroscope, the linear and rotational
kinematics of the aircraft in take-off, landing and yaw can be
analysed. To get a deeper insight, the acceleration values were
numerically integrated using a simple scheme to obtain the vertical
speed and height.  The measured maximum value of the vertical speed
can be compared with the manufacturer's specifications.

\section{The outdoor experiment}

\begin{figure}[h]
\begin{center}
\includegraphics[width=0.6\textwidth]{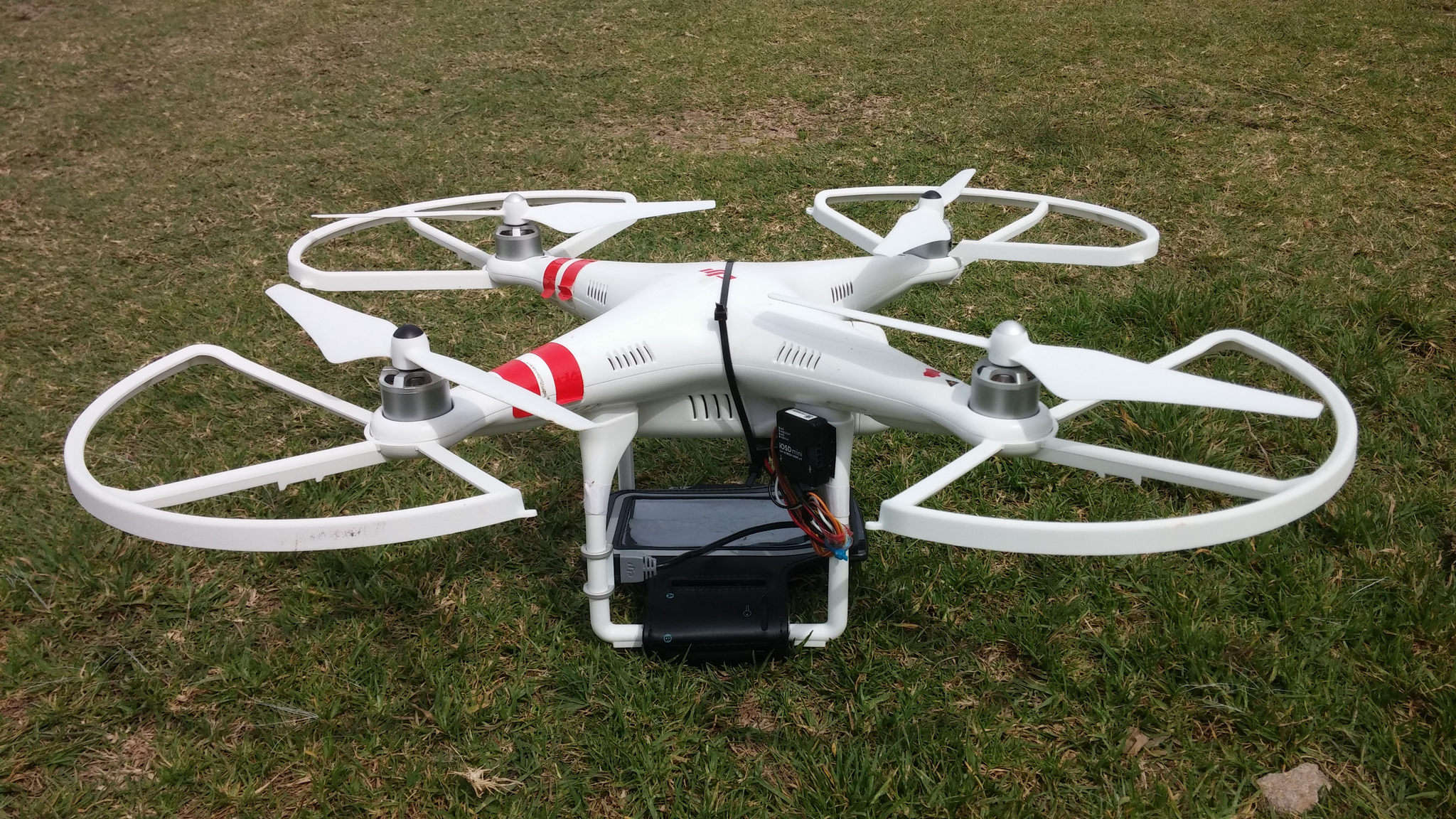}
\caption{Smartphone mounted on DJI Phantom 2 using an armband.}
\label{fig1}
\end{center}
\end{figure}

A smartphone, model LG G2, was mounted on a radio-controlled a DJI
Phantom 2 quadcopter using an armband case as shown in
Fig.~\ref{fig1}. Several simple flights were accomplished consisting
of ascents and descents with constant speed or constant acceleration
and different trajectories at constant height (circles, “U” or “S”
shaped). During these flights, the smartphone's sensors recorded the
acceleration and the angular velocity using the Androsensor
application. In practice, after several minutes of collecting data, it
was difficult to discern which portion of the graph corresponds to a
particular manoeuvre. To avoid this problem a video camera, or a
second smartphone, was used from the ground to record alternatively
the flight and the operation of the remote control. In what follows,
we focus on some of the records of the sensors and analyse the
acceleration and angular velocity along the vertical axis.

\section{The drone's flight}

The drone's flight is based on the rotation of four vertical axis
propellers (\textit{rotors}) arranged in the corners of a square. As a
result of the Newton's second law, when the drone is hovering at
constant altitude the upwards thrust generated by the rotors equals
the downward gravitational force on the airframe.  A vertical
acceleration or deceleration is obtained thanks to an increase or
decrease of the total power of the rotors.  If the thrust force is
upwards and exceeds the force of gravity, then the drone will
initially accelerate, however, in doing so its potential energy
increases, and the motors have to do work to supply that energy.  So,
the rate of rise will be limited to the rate at which the motors can
deliver this energy.

\begin{figure}[th]
\begin{center}
\includegraphics[width=0.6\textwidth]{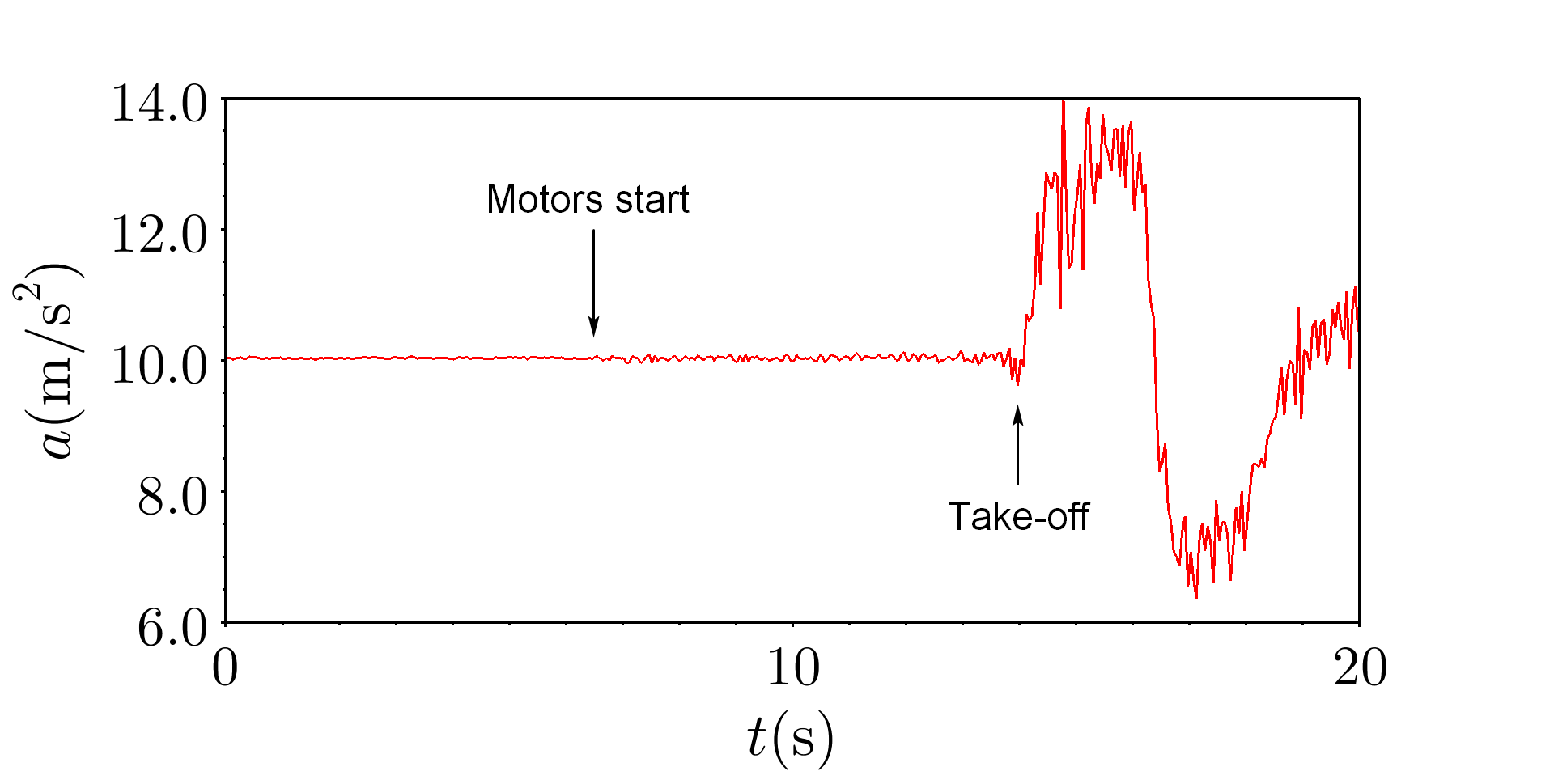}
\includegraphics[width=0.6\textwidth]{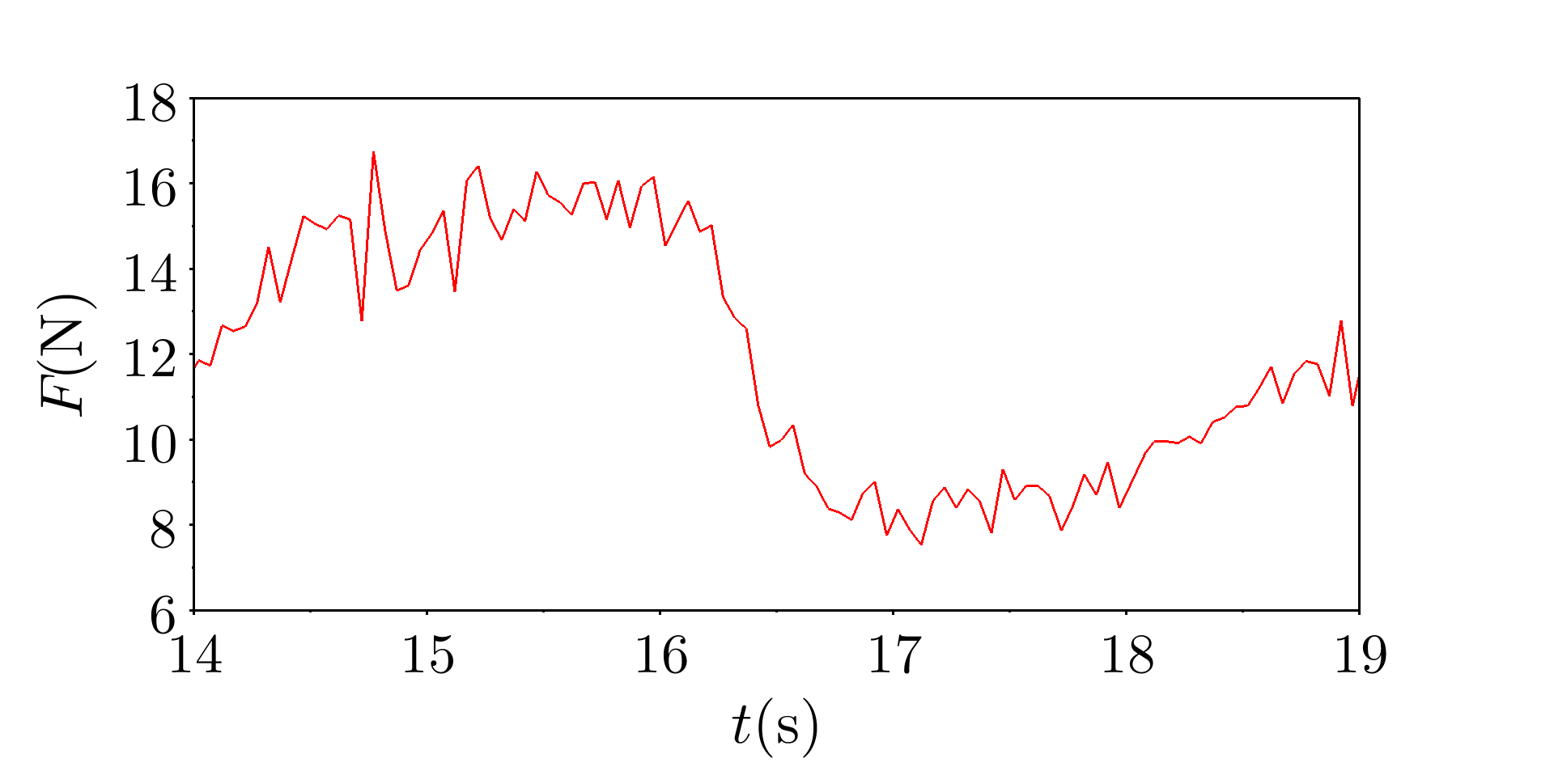}
\caption{Record of the acceleration sensor (top) and force (bottom)
  during the take-off. In all the cases the sampling frequency is set
  to 20 samples per second.}
\label{fig2}
\end{center}
\end{figure}

The vertical acceleration as a function of time is shown in
Fig.~\ref{fig2}.  Initially the drone is turned off on the
ground. After a few seconds, the vehicle is switched on and the
propellers start to spin. The vibrations induced by the propellers are
clearly detected by the accelerometer. Then, the drone takes off as
revealed by an upward acceleration.  It is also worth mentioning that
the vertical acceleration does not reach its maximum value inmediatly
but it takes an interval of about 1 s before that.  After an interval
of deceleration, the drone is kept hovering for a few seconds.  Note
that the acceleration values registered by the sensors are not
centered about $0$ but about 10 m/s$^2$. In the accelerometer
essentially the force required to hold a mass in a fixed position
relative to the smartphone is being registered. In accordance with
Newton's second law this force is proportional to the acceleration. At
the same time, quite independently, the mass is being acted on by a
completely different force of 10 N/kg downwards due to the mutual
gravitational attraction of the mass in the instrument and the mass of
the Earth. Because the gravity force per unit mass cannot be
distinguished from the acceleration force in this instrument (or
indeed by anyone blindfolded in an aircraft), it appears as a
\textit{fictitious} acceleration (of 10 m/s$^2$, which is exactly the
same thing as 10 N/kg) in the direction away from the centre of the
Earth. This fictitious acceleration also appears in other smartphone
experiments as the physical pendulum
\cite{monteiro2014exploring,MONTEIRO2015}. It is also worth noting
that the registered vertical acceleration is greater than 10 m/s$^2$
at the take off. Nevertheless, a few seconds later when the device
decelerate it is lesser than 10 m/s$^2$.

\begin{figure}[h]
\begin{center}
\includegraphics[width=0.8\textwidth]{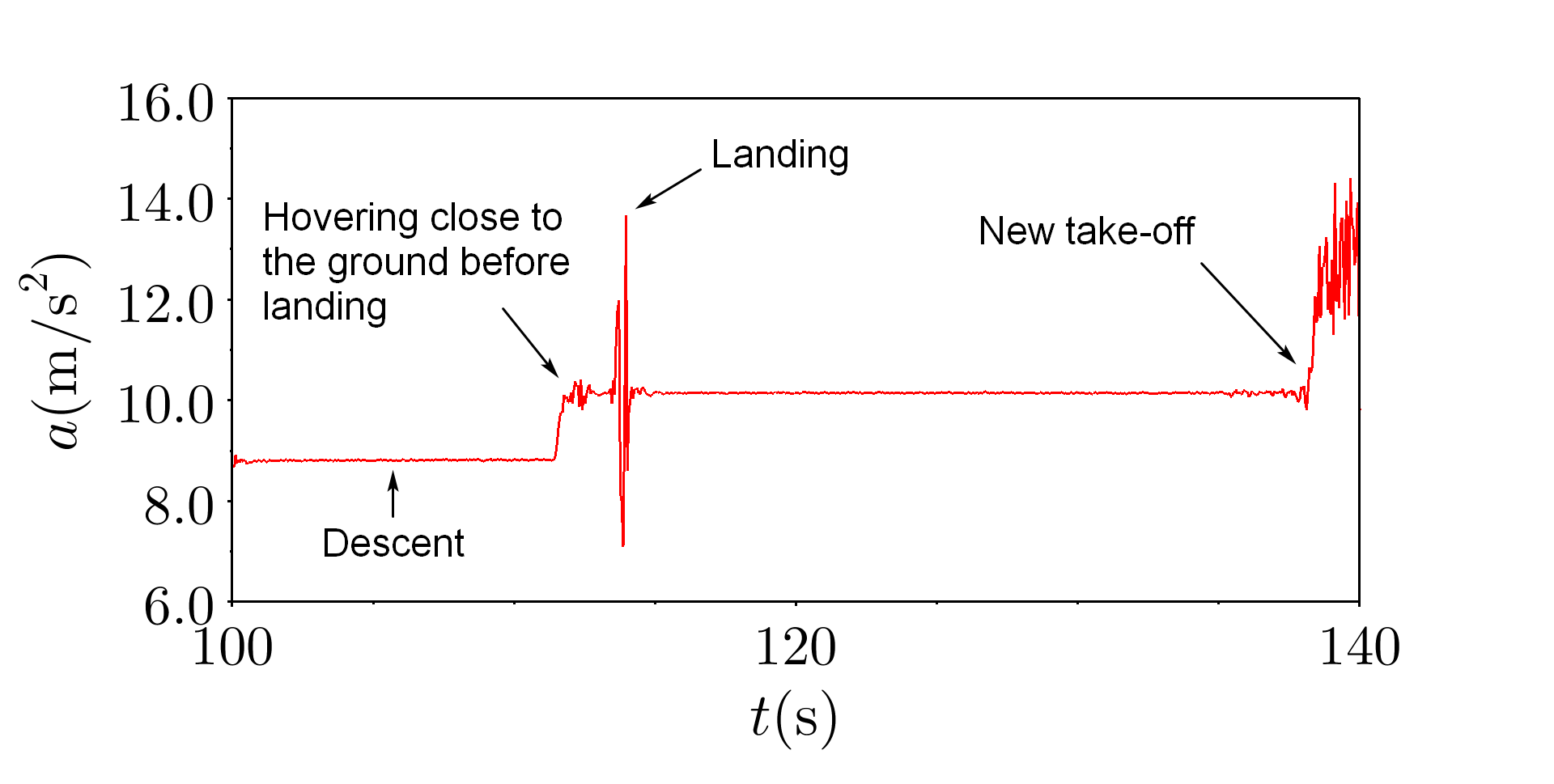}
\caption{Vertical acceleration during a landing and a subsequent take off.}
\label{fig3}
\end{center}
\end{figure}

The landing and a subsequent take off of the drone is analyzed in
Fig.~\ref{fig3}.  During the first stage the device is accelerating
downward, then the acceleration registered is \textit{less} than 10
m/s$^2$. To avoid an abrupt landing, the drone is kept hovering close
to the ground and smoothly landed at $t=174$ s. After about 20 s a new
take off is accomplished. As a consequence of the stabilization
mechanism of the drone when it is ascending or descending at constant
speed or hovering the acceleration measurements are very noisy and
special care should be taken to numerically integrate the acceleration
to obtain the velocity or the altitude \cite{monteiro2016using}.  The
origin of the noise can be associated to several sources. On the one
hand, there are internal contributions from the motors evidenced in the
increase of the noise at the starting time indicated in the figure. On the other hand, when the drone takes off its stabilization system responds to the
atmospheric conditions leading to an additional rise in the vibrations
registered by the smartphone acceleromenter.

Generally speaking, the design of a drone is based on the rotational
dynamics of the rotors and involves subtle aspects. When the rotation
axis is parallel to the spin axis, torque is generated in the airframe
by various effects, in particular, due to rotational inertia and also
as a reaction to the force needed to overcome the air resistance and
to push it downwards.  The ability to deliver power to each rotor
indenpendently provides the possibility to control the drone. As a
consequence of the conservation of angular momentum, to maintain the
angular velocity of the device, the sum of the torques generated by
the four rotors must be zero.  If it is not the case, the device will
undergo accelerating vertical-axis rotation. In particular, yaw
(rotation about the vertical axis of the machine) is obtained
increasing the power in one pair of diagonally opposite rotors and
decreasing the power in the other pair to maintain constant lift. The
horizontal flight is based on a similar mechanism. To fly forward or
backward a horizontal component of thrust is achieved by increasing
the rotation rate of two adyacent rotors (the rear ones) and a
decreasing of the other two (the front ones). In this situation, the
greater force in the back leads the drone to tilt forward, resulting
in the thrust force having a horizontal component.

\begin{figure}[h]
\begin{center}
\includegraphics[width=0.8\textwidth]{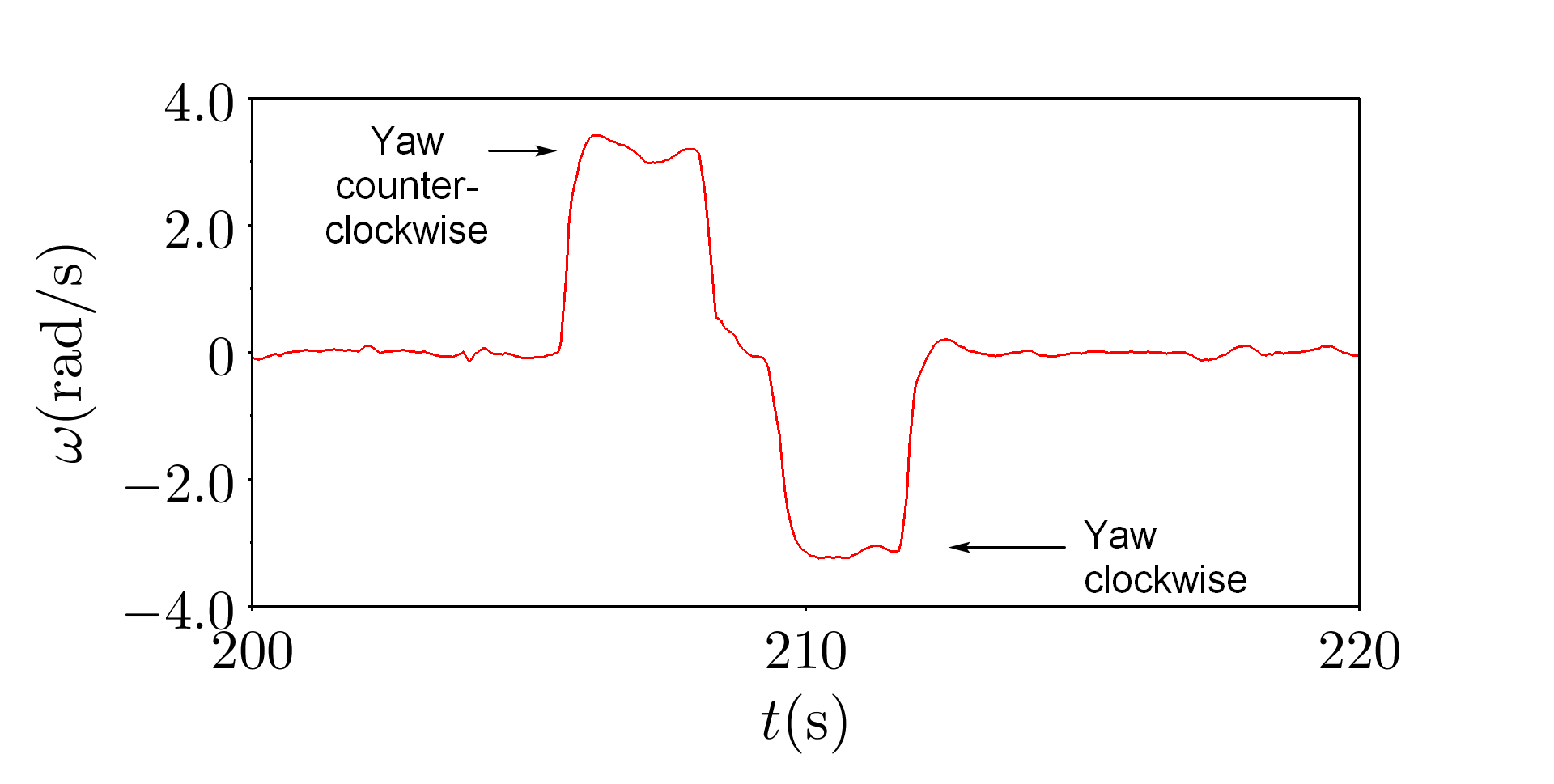}
\includegraphics[width=0.8\textwidth]{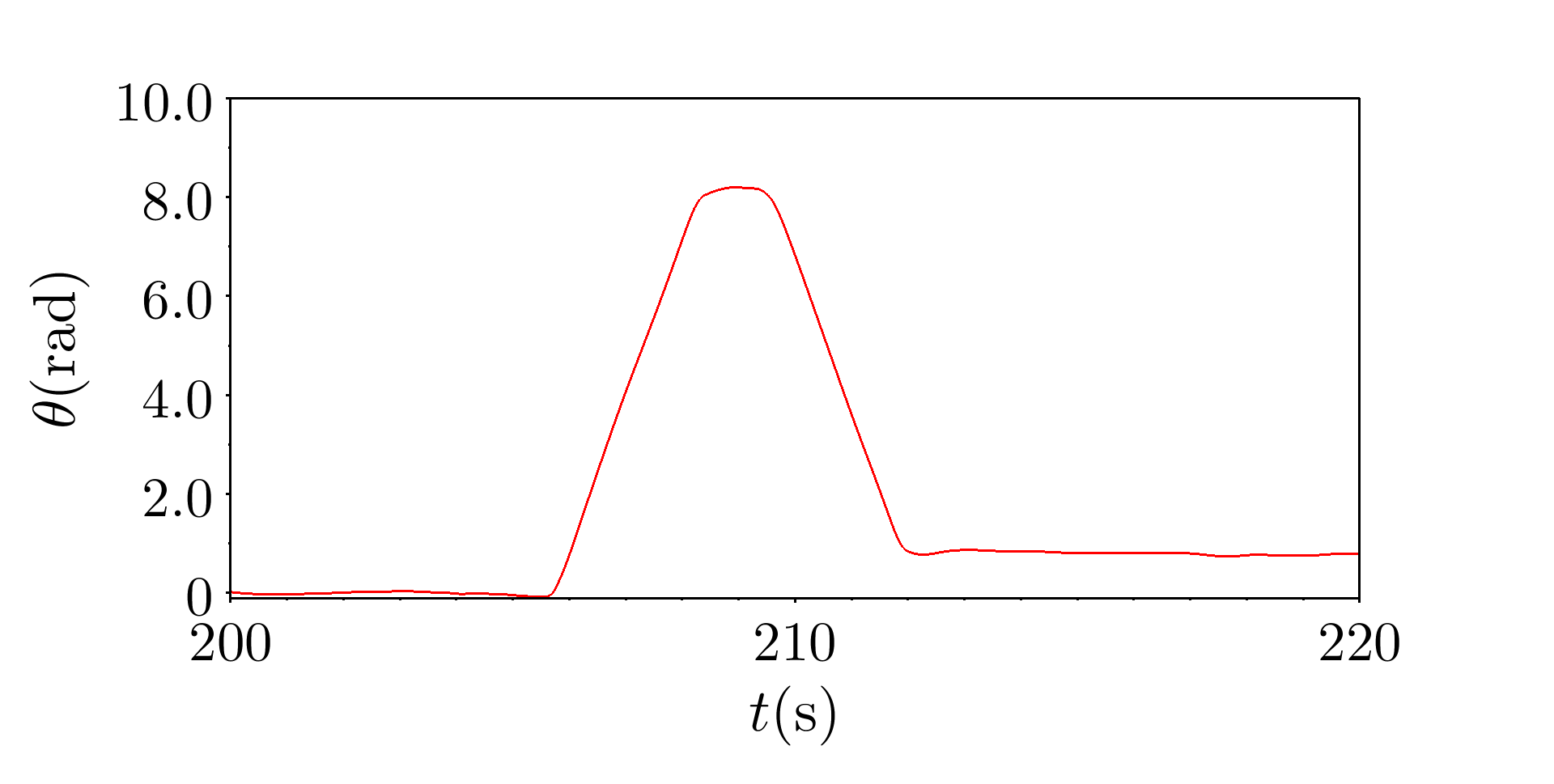}
\caption{Angular velocity (vertical component) recorded by the gyroscope
(top panel). Angle of rotation in a horizontal plane obtained by the numerical
integration of the vertical angular velocity or yaw (bottom panel).}
\label{fig4}
\end{center}
\end{figure}

The results obtained by the rotation sensor are depicted in
Fig.~\ref{fig4}. Yaw, both clockwise and counter-clockwise (see from
above), is easily distinguished in the vertical component of the
angular velocity. The maximum values attained by the angular velocity
can be compared with that provided by the manufacturer. In our model
the technical specification indicates $200\degree/$s $\sim 3.49$ rad/s
which agrees very well the maximum values observed in the figure.

\section{Vertial speed, heigth and angle}

\begin{figure}[h]
\begin{center}
\includegraphics[width=0.9\textwidth]{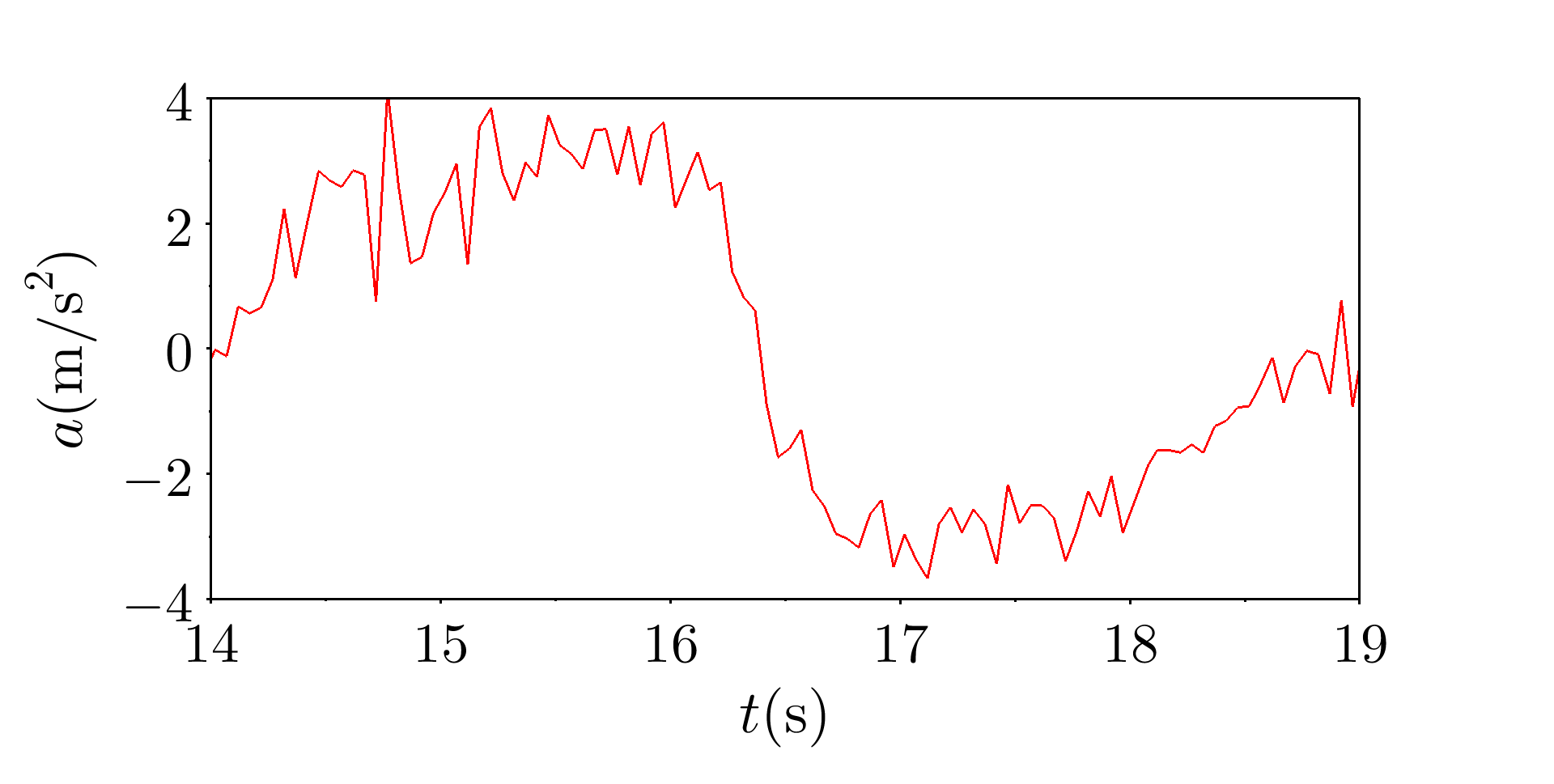}
\includegraphics[width=0.4\textwidth]{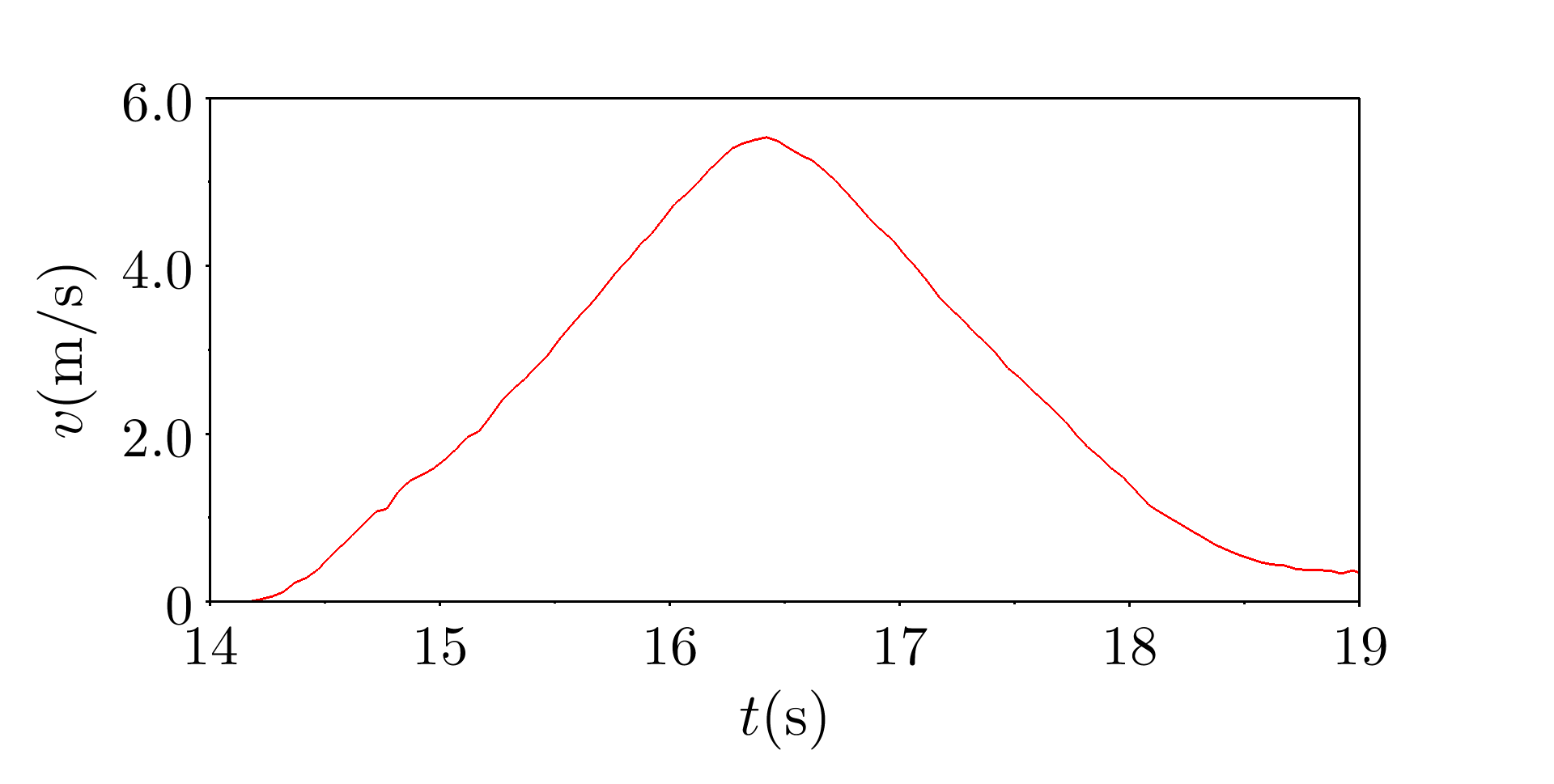}
\includegraphics[width=0.4\textwidth]{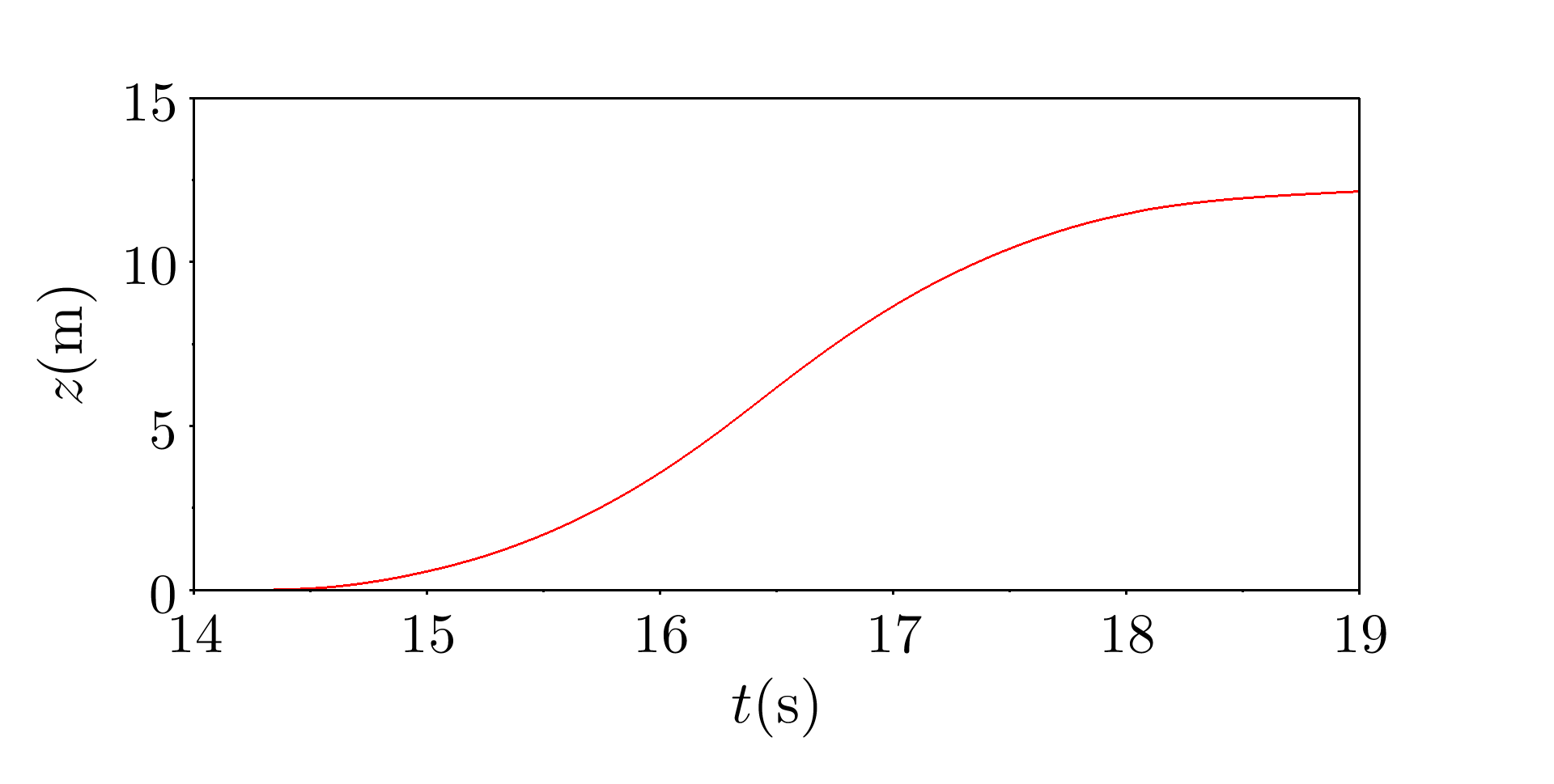}
\caption{(top panel) Vertical acceleration (after correction for
  gravitational force per unit mass), (bottom left) vertical component
  of velocity, (bottom right) vertical displacement obtained by
  numerical integration of the sensor data.}
\label{fig5}
\end{center}
\end{figure}

Using the vertical component of the acceleration it is possible to
integrate numerically these value to obtain, firstly, the vertical
speed, and, then, the height as shown in Fig.~\ref{fig5}. This
procedure is similar to those proposed in other contexts
\cite{monteiro2014exploring,Vogt2014analyzing,Monteiro2017magnetic}.
As the drone  speed control was set at maximum during a time interval the
vertical speed attained was the maximum possible for this device
model. In our case, the maximum ascent speed reported by the
manufacturer was 6 m/s which agrees very well with that obtained
numerically as it can be seen in the figure. In case of having a large
and clear space it is also possible to supplement these results with
video analysis \cite{Monteiro2021allies}.

Another exercise consists in numerically integrating the vertical
angular velocity to obtain the rotation angle in a horizontal
plane. The numerical result is shown in Fig.~\ref{fig4} (bottom). Notice that
this can be easily done when the rotation is uniaxial, however, when
the drone is rotating along two or three axis this task is very
difficult.

\section{Conclusion}

This experiment, which can be easily replicated in a school yard,
provides an interesting opportunity to understand basic concepts in
aerodynamics by using the smartphone's sensors and analyzing their
results.  The relationships between various kinematic quantities, such
as the acceleration, velocity, displacement, angular velocity, and
angle, can be further investigated using numerical methods. With this
simple experiment we exploit the educational capabilities provided by
two expanding technologies: drones and smartphones.

\section*{Acknowledgements} We acknowledge financial support from grant
Fisica Nolineal (ID 722) Programa Grupos I + D CSIC 2018 (UdelaR,
Uruguay) and PEDECIBA (UdelaR, MEC, Uruguay). We thank the anonymous
reviewer for his/her critical reading and valuble suggestions.

\section*{References}

\providecommand{\newblock}{}


\begin{thebibliography}{10}
\expandafter\ifx\csname url\endcsname\relax
  \def\url#1{{\tt #1}}\fi
\expandafter\ifx\csname urlprefix\endcsname\relax\def\urlprefix{URL }\fi
\providecommand{\eprint}[2][]{\url{#2}}

\bibitem{floreano2015science}
Floreano D and Wood R~J 2015 {\em Nature\/} {\bf 521} 460--466

\bibitem{Obaid_2021}
Obaid M~S and Mebayet S~O 2021 {\em Journal of Physics: Conference Series\/}
  {\bf 1818} 012104
  \urlprefix\url{https://doi.org/10.1088/1742-6596/1818/1/012104}

\bibitem{saifizi2021using}
Saifizi M, Syahirah N, Mustafa W~A, Rahim H~A and Nasrudin M~W 2021 Using
  unmanned aerial vehicle in 3d modelling of uniciti campus to estimate
  building size {\em Journal of Physics: Conference Series\/} vol 1962 (IOP
  Publishing) p 012057

\bibitem{monteiro2016exploring}
Monteiro M, Vogt P, Stari C, Cabeza C and Marti A~C 2016 {\em The Physics
  Teacher\/} {\bf 54} 308--309

\bibitem{monteiro2014exploring}
Monteiro M, Cabeza C and Mart{\'\i} A~C 2014 {\em European Journal of
  Physics\/} {\bf 35} 045013

\bibitem{MONTEIRO2015}
Monteiro M, Cabeza C and Marti A~C 2015 {\em Revista Brasileira de Ensino de
  F\'i\-sica\/} {\bf 37} 1303 -- ISSN 1806-1117

\bibitem{monteiro2016using}
Monteiro M and Marti A~C 2017 {\em Physics Education\/} {\bf 52} 015010

\bibitem{Vogt2014analyzing}
Vogt P and Kuhn J 2014 {\em The Physics Teacher\/} {\bf 52} 118--119

\bibitem{Monteiro2017magnetic}
Monteiro M, Stari C, Cabeza C and Marti A~C 2017 {\em The Physics Teacher\/}
  {\bf 55} 580--581

\bibitem{Monteiro2021allies}
Monteiro M, Cabeza C, Stari C and Marti A~C 2021 {\em Journal of Physics:
  Conference Series\/} {\bf 1929} 012038
  \urlprefix\url{https://doi.org/10.1088/1742-6596/1929/1/012038}

\end{thebibliography}
\end{document}